\documentclass[final]{svjour3}
\usepackage{graphicx}
\usepackage{rotating}
\usepackage{amssymb}
\usepackage{mathptmx}
\usepackage[numbers,sort&compress]{natbib}
\usepackage{hyperref}
\makeatletter
\usepackage{xspace}

\newcommand{\spider}{\textsc{Spider}\xspace}
\newcommand{\planck}{\textit{Planck}\xspace}

\def\eg{{\it e.g.}\xspace}

\def\spider{{\sc Spider}\xspace}

\bibpunct[,]{[}{]}{,}{n}{,}{,}

\usepackage[usenames]{color}

\begin{document}

\newcommand{\hdblarrow}{H\makebox[0.9ex][l]{$\downdownarrows$}-}
\title{In-Flight Gain Monitoring of SPIDER's Transition-Edge Sensor Arrays}

\author{
J.~P.~Filippini$^{1,2}$ \and A.~E.~Gambrel$^{3}$ \and A.~S.~Rahlin$^{4,3}$ \and E.~Y.~Young$^{5,6}$ \and P.~A.~R.~Ade$^{7}$ \and M.~Amiri$^{8}$ \and S.~J.~Benton$^{9}$ \and A.~S.~Bergman$^{9}$ \and R.~Bihary$^{10}$ \and J.~J.~Bock$^{11,12}$ \and J.~R.~Bond$^{13}$ \and J.~A.~Bonetti$^{12}$ \and S.~A.~Bryan$^{14}$ \and H.~C.~Chiang$^{15,16}$ \and C.~R.~Contaldi$^{17}$ \and O.~Dor{\'e}$^{11,12}$ \and A.~J.~Duivenvoorden$^{9,18}$ \and H.~K.~Eriksen$^{19}$ \and M.~Farhang$^{20,13,21}$ \and A.~A.~Fraisse$^{9}$ \and K.~Freese$^{22,18}$ \and M.~Galloway$^{19}$ \and N.~N.~Gandilo$^{23}$ \and K.~Ganga$^{24}$ \and R.~Gualtieri$^{25}$ \and J.~E.~Gudmundsson$^{18}$ \and M.~Halpern$^{8}$ \and J.~Hartley$^{26}$ \and M.~Hasselfield$^{27}$ \and G.~Hilton$^{28}$ \and W.~Holmes$^{12}$ \and V.~V.~Hristov$^{11}$ \and Z.~Huang$^{13}$ \and K.~D.~Irwin$^{5,6}$ \and W.~C.~Jones$^{9}$ \and A.~Karakci$^{19}$ \and C.~L.~Kuo$^{5}$ \and Z.~D.~Kermish$^{9}$ \and J.~S.-Y.~Leung$^{21,29}$ \and S.~Li$^{9,30}$ \and D.~S.~Y.~Mak$^{17}$ \and P.~V.~Mason$^{11}$ \and K.~Megerian$^{12}$ \and L.~Moncelsi$^{11}$ \and T.~A.~Morford$^{11}$ \and J.~M.~Nagy$^{31,32}$ \and C.~B.~Netterfield$^{21,26}$ \and M.~Nolta$^{13}$ \and R. O\rq Brient$^{12}$ \and B.~Osherson$^{1}$ \and I.~L.~Padilla$^{21,33}$ \and B.~Racine$^{19}$ \and C.~Reintsema$^{28}$ \and J.~E.~Ruhl$^{10}$ \and M.~C.~Runyan$^{11}$ \and T.~M.~Ruud$^{19}$ \and J.~A.~Shariff$^{13}$ \and E.~C.~Shaw$^{1}$ \and C.~Shiu$^{9}$ \and J.~D.~Soler$^{34}$ \and X.~Song$^{9}$ \and A.~Trangsrud$^{11,12}$ \and C.~Tucker$^{7}$ \and R.~S.~Tucker$^{11}$ \and A.~D.~Turner$^{12}$ \and J.~F.~van~der~List$^{9}$ \and A.~C.~Weber$^{12}$ \and I.~K.~Wehus$^{19}$ \and S.~Wen$^{10}$ \and D.~V.~Wiebe$^{8}$
}

\institute{
$^{1}$Department of Physics, University of Illinois at Urbana-Champaign, Urbana, IL, USA\\
$^{2}$Department of Astronomy, University of Illinois at Urbana-Champaign, Urbana, IL, USA\\
$^{3}$Kavli Institute for Cosmological Physics, University of Chicago, Chicago, IL USA\\
$^{4}$Fermi National Accelerator Laboratory, Batavia, IL, USA\\
$^{5}$Department of Physics, Stanford University, Stanford, CA , USA\\
$^{6}$SLAC National Accelerator Laboratory, Menlo Park, CA, USA\\
$^{7}$School of Physics and Astronomy, Cardiff University, Cardiff, UK\\
$^{8}$Department of Physics and Astronomy, University of British Columbia, Vancouver, BC, Canada\\
$^{9}$Department of Physics, Princeton University, Princeton, NJ, USA\\
$^{10}$Physics Department, Case Western Reserve University, Cleveland, OH, USA\\
$^{11}$Division of Physics, Mathematics and Astronomy, California Institute of Technology, Pasadena, CA, USA\\
$^{12}$Jet Propulsion Laboratory, Pasadena, CA , USA\\
$^{13}$Canadian Institute for Theoretical Astrophysics, University of Toronto, Toronto, ON, Canada\\
$^{14}$School of Electrical, Computer, and Energy Engineering, Arizona State University, Tempe, AZ, USA\\
$^{15}$Department of Physics, McGill University, Montreal, QC, Canada\\
$^{16}$School of Mathematics, Statistics and Computer Science, University of KwaZulu-Natal, Durban, South Africa\\
$^{17}$Blackett Laboratory, Imperial College London, London, UK\\
$^{18}$The Oskar Klein Centre for Cosmoparticle Physics, Department of Physics, Stockholm University, Stockholm, Sweden\\
$^{19}$Institute of Theoretical Astrophysics, University of Oslo, Oslo, Norway\\
$^{20}$Department of Physics, Shahid Beheshti University, Tehran Iran\\
$^{21}$Department of Astronomy and Astrophysics, University of Toronto, Toronto, ON, Canada\\
$^{22}$Department of Physics, University of Texas, Austin, TX, USA\\
$^{23}$Steward Observatory, Tucson, AZ, USA\\
$^{24}$APC, Univ. Paris Diderot, CNRS/IN2P3, CEA/Irfu, Obs de Paris, Sorbonne Paris Cit\'e, France\\
$^{25}$High Energy Physics Division, Argonne National Laboratory, Argonne, IL, USA\\
$^{26}$Department of Physics, University of Toronto, Toronto, ON, Canada\\
$^{27}$Department of Astronomy and Astrophysics, Pennsylvania State University, University Park, PA, USA\\
$^{28}$National Institute of Standards and Technology, Boulder, CO, USA\\
$^{29}$Dunlap Institute for Astronomy and Astrophysics, University of Toronto, Toronto, ON, Canada\\
$^{30}$Department of Mechanical and Aerospace Engineering, Princeton University, Princeton, NJ, USA\\
$^{31}$Department of Physics, Washington University in St. Louis, St.  Louis, MO, USA\\
$^{32}$McDonnell Center for the Space Sciences, Washington University in St. Louis, St.  Louis, MO, USA\\
$^{33}$Department of Physics and Astronomy, Johns Hopkins University, Baltimore, MD, USA\\
$^{34}$Max-Planck-Institute for Astronomy, Heidelberg, Germany\\
\email{jpf@illinois.edu}
}
\authorrunning{J.P.~Filippini \textit{et. al}}

\maketitle

\begin{abstract}

Experiments deploying large arrays of transition-edge sensors (TESs) often require a robust method to monitor gain variations with minimal loss of observing time. We propose a sensitive and non-intrusive method for monitoring variations in TES responsivity using small square waves applied to the TES bias. We construct an estimator for a TES's small-signal power response from its electrical response that is exact in the limit of strong electrothermal feedback. We discuss the application and validation of this method using flight data from SPIDER, a balloon-borne telescope that observes the polarization of the cosmic microwave background with more than 2000 TESs. This method may prove useful for future balloon- and space-based instruments, where observing time and ground control bandwidth are limited. 


\end{abstract}

\section{Introduction}
Modern instruments to observe the cosmic microwave background (CMB)~\cite{pdg} typically employ large arrays of 
superconducting transition-edge sensor (TES) bolometers~\cite{IrwinHilton2005} 
to detect its faint variations in intensity and polarization across the sky. 
Due to the ``self-calibrating'' nature of electrothermal feedback, it is straightforward to convert the 
recorded data stream (varying TES current) into an estimate of the variation in power absorbed by the associated bolometer, and from there into sky brightness.
The precision of this estimate is limited by the accuracy of bolometer and instrument modeling, however, and is generally not sufficient for absolute calibration of the instrument.
Overall calibration typically uses the CMB itself as a reference source, using observations accumulated over sufficiently long observing time scales (often groups of detectors over many sky scans).
Experiments use various strategies to track relative variations in the response of single detectors over shorter time scales, 
including calibration lamps~\cite{Piacentini2001} and atmospheric brightness variations~\cite{Takahashi2008}.

\spider~\cite{Fraisse2013,Spider2021} is a balloon-borne instrument designed to measure the polarization of the CMB from an altitude of $\sim$36~km.
This altitude provides a pristine view of the millimeter-wave sky, with minimal contamination by atmospheric emission.
\spider's first flight in 2015 observed the sky at 95 and 150~GHz with more than 2000 antenna-coupled TESs~\cite{JPLdet2015}, setting constraints on primordial gravitational waves~\cite{Spider2021} and circular polarization~\cite{SpiderVpol}.
While \spider's overall maps are calibrated against \planck temperature maps at 100 and 143~GHz~\cite{planck18_hfi}, 
the signal-to-noise is insufficient for calibration on the time scale of a few scans.
\spider's architecture and observing strategy complicate other common methods for short-term gain monitoring:
atmospheric emission is minimal, the refractive optics provide no natural location for a calibration lamp, 
and calibration from astrophysical sources (\eg, planets) was impractical due to \spider's relatively large ($\sim30'$) beams and the limited sky accessible during the polar day.

In this paper we describe a technique for monitoring gain variations in TES systems using small changes in TES bias current (``bias steps''). 
Our method is simple and non-disruptive, requiring no dedicated hardware and consuming negligible observing time.
This technique was developed for the \spider experiment as part of pre-flight testing in 2011 and implemented for its January 2015 flight and subsequent analysis.
It primarily targets the possible effects on TES responsivity from small changes in the TES power budget (optical loading, wafer temperature) over hours/days as \spider changes location/altitude and its helium tank drains; other sources of gain drift are possible, notably from magnetic fields (expected to be negligible due to extensive shielding~\cite{Runyan2010}) and TES bias drifts.
Other teams have described applications of bias steps to characterize similar TES arrays, 
but their published use has generally been limited to estimates of device resistance~\cite{Pappas2016}, time constants~\cite{Stevens2020}, or complex impedance~\cite{Cothard2020}, rather than real-time gain monitoring.
We discuss theoretical background in Sect.~\ref{sec:theory}, in-flight performance in Sect.~\ref{sec:flight}, and conclude in Sect.~\ref{sec:conclusion}.

\section{Optical and Electrical TES Response}
\label{sec:theory}

\subsection{Low-Frequency TES Response}
In this work, we consider the response of a simple TES to small perturbations in its power balance and electrical bias in the low-frequency limit.
We consider a TES with superconducting transition temperature $T_c$ and resistance $R$ wired in parallel to a shunt resistance $R_{sh}\ll R$, the pair driven by a bias current $I_b$ to yield a voltage bias of the TES itself.
The TES is fabricated onto a suspended island of heat capacity, $C$, which is isolated from a thermal bath at $T_b$ by a thermal conductance $G$.
The linear theory of such a TES has been summarized by Irwin and Hilton \cite{IrwinHilton2005}, 
who derive expressions for these small-signal responses as functions of frequency ($\omega=2\pi f$).
For our purposes we limit our discussion to the low-frequency ($\omega\to 0$) limits of these expressions, 
as we consider bias steps and sky signals that are slow compared to the electrothermal feedback response of \spider's TESs.

Following Irwin and Hilton (Equation~37), we can write the low-frequency response of the measured TES current, $I$, to a small perturbation in the power $P$ incident on the island as
\begin{equation}
s_0 \equiv \frac{dI}{dP}(\omega=0) = -\frac{1}{I R} \left( \left( 1 - \frac{R_{sh}}{R} \right) + \frac{1+\beta_I+\frac{R_{sh}}{R}}{\mathcal{L}_I} \right)^{-1}.
\label{eqn:sI_0}
\end{equation}
Here, $\mathcal{L}_I\equiv \frac{\alpha_I I^2 R}{G T_c}$ is the dimensionless electrothermal loop gain, while 
$\alpha_I \equiv \frac{d \log R}{d \log T}$ and $\beta_I \equiv \frac{d \log R}{d \log I}$
are the TES's dimensionless responsivities to changes in temperature and current, respectively.
Note that the power responsivity is negative, in that a small increase in power appears as a small decrease in TES current.

Similarly, the electrical response of the TES circuit to small changes in the bias current can be derived using 
the TES complex impedance (Irwin and Hilton, Equation~42):
\begin{equation}
Z_{TES}(\omega) = R(1+\beta_I) + \frac{R \mathcal{L}_I}{1-\mathcal{L}_I}\frac{2+\beta_I}{1+i\omega \tau_I},
\label{eqn:Ztes}
\end{equation}
where $\tau_I\equiv \frac{C/G}{1-\mathcal{L}_I}$.
Considering this impedance in parallel with $R_{sh}$ and taking $\omega\to 0$, one can derive the observable current response to small changes in the bias current $I_b$:
\begin{equation}
\eta_0 \equiv \frac{d I}{d I_b}(\omega=0) = - \frac{R_{sh}}{R} \left( \left(1 - \frac{R_{sh}}{R}\right) + \frac{2+\beta_I}{\mathcal{L}_I-1} \right)^{-1}.
\label{eqn:eta_0}
\end{equation}
Note that the electrical responsivity ($\eta_0$) as defined here is dimensionless, while the power responsivity ($s_0$) has the dimension of inverse voltage.

\subsection{Gain estimator}

In the limit of high loop gain $(\mathcal{L}_I \gg 1)$, Eqs.~\ref{eqn:sI_0} and \ref{eqn:eta_0} take very simple forms:
\begin{equation}
\lim_{\mathcal{L}_I\to\infty} s_0 = -\frac{1}{I_b R_{sh}} \left( \frac{1+R_{sh}/R}{1-R_{sh}/R} \right),
\label{eqn:s_0_limit}
\end{equation}
\begin{equation}
\lim_{\mathcal{L}_I\to\infty} \eta_0 = -\frac{R_{sh}/R}{1-R_{sh}/R}.
\label{eqn:eta_0_limit}
\end{equation}
Note the substitution $I = \frac{I_b R_{sh}}{R+R_{sh}}$ in deriving Eq.~\ref{eqn:s_0_limit} from Eq.~\ref{eqn:sI_0}. 
The second of these expressions is particularly useful, as it relates the measured bias step amplitude to the TES resistance.
\spider's in-flight monitoring system uses this to regulate the TES bias point.

Equations~\ref{eqn:s_0_limit} and \ref{eqn:eta_0_limit} suggest a simple estimator for the power response in terms of the electrical response:
\begin{equation}
\sigma_0 \equiv \frac{2 \eta_0 - 1}{I_b R_{sh}} = s_0 + \mathcal{O}\left(\frac{1}{\mathcal{L}_I}\right)
\label{eqn:si0est}
\end{equation}
The equality between $\sigma_0$ and $s_0$ is exact when $\mathcal{L}_I\gg1$, and thus is likely to be a useful approximation for $\mathcal{L}_I$ larger than a few.

\subsection{Simulation results}
To evaluate the performance of the $s_0$ estimator, we performed a series of numerical simulations.
We model a \spider bolometer as a single heat capacity, $C\sim 0.9$~pJ/K, isolated from a thermal bath at $T_b=0.3$~K by a thermal conductance $G\sim20$~pW/K.
The heat capacity supports a TES with normal resistance $R_n=32$~m$\Omega$ and transition temperature $T_c=0.5$~K, wired in parallel with a shunt resistor $R_{sh}=3$~m$\Omega$.
We assume a $\tanh$ model for the transition~\cite{Burney2006}, with a loop gain of $\mathcal{L}_I \sim 20$ at $R\sim 0.5 R_n$, 
consistent with \spider's relatively low optical loading in flight ($\sim0.3$~pW absorbed power).
We generate libraries of simulated TES load curves ($I$ as a function of $I_b$) under different absorbed powers.
By slicing through this data at fixed bias current, we simulate the change in TES parameters and gains due to changes in power from any given initial bias point.

\begin{figure}
  \centering
  \includegraphics[width=0.49\linewidth, keepaspectratio]{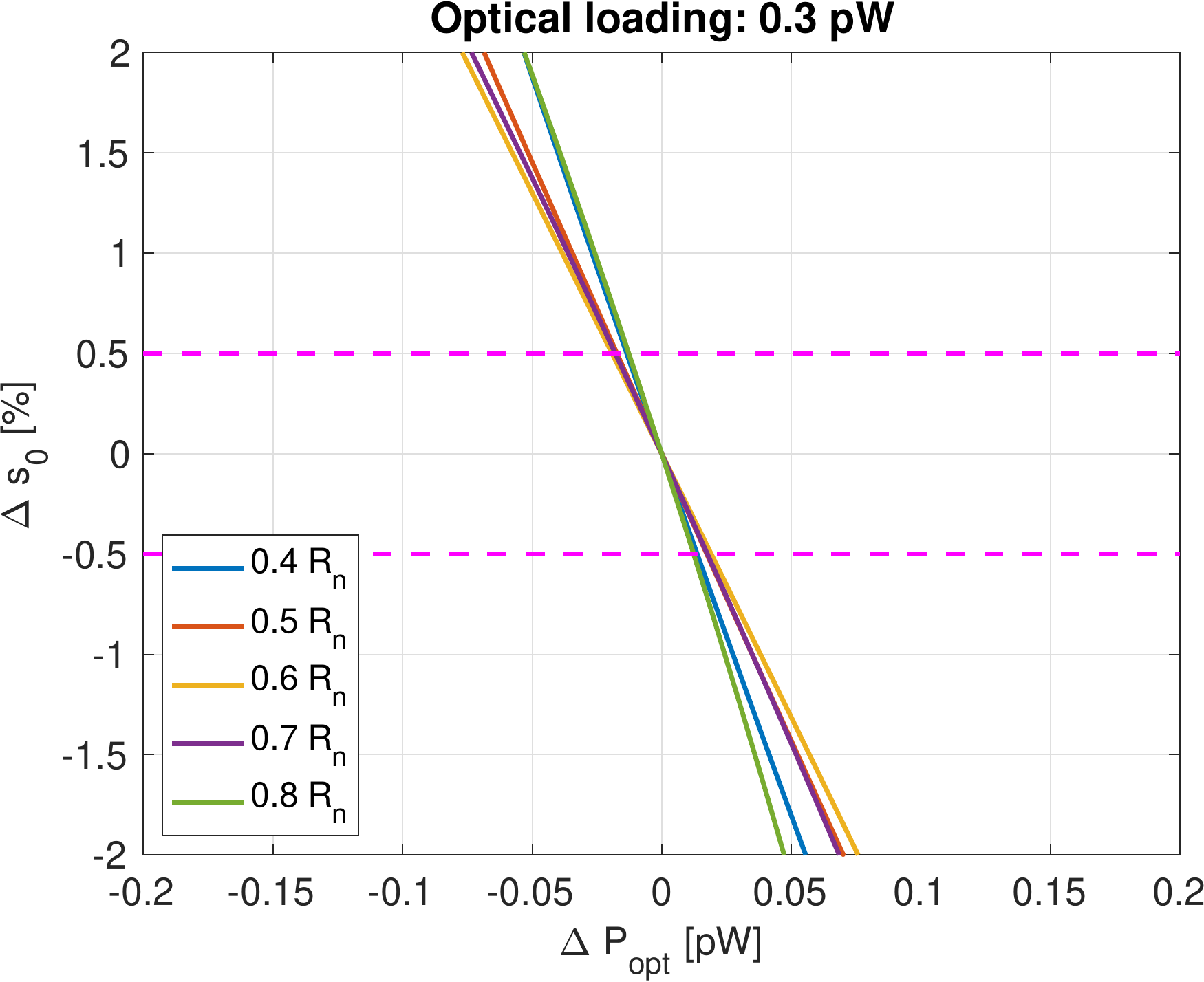}
  \includegraphics[width=0.49\linewidth, keepaspectratio]{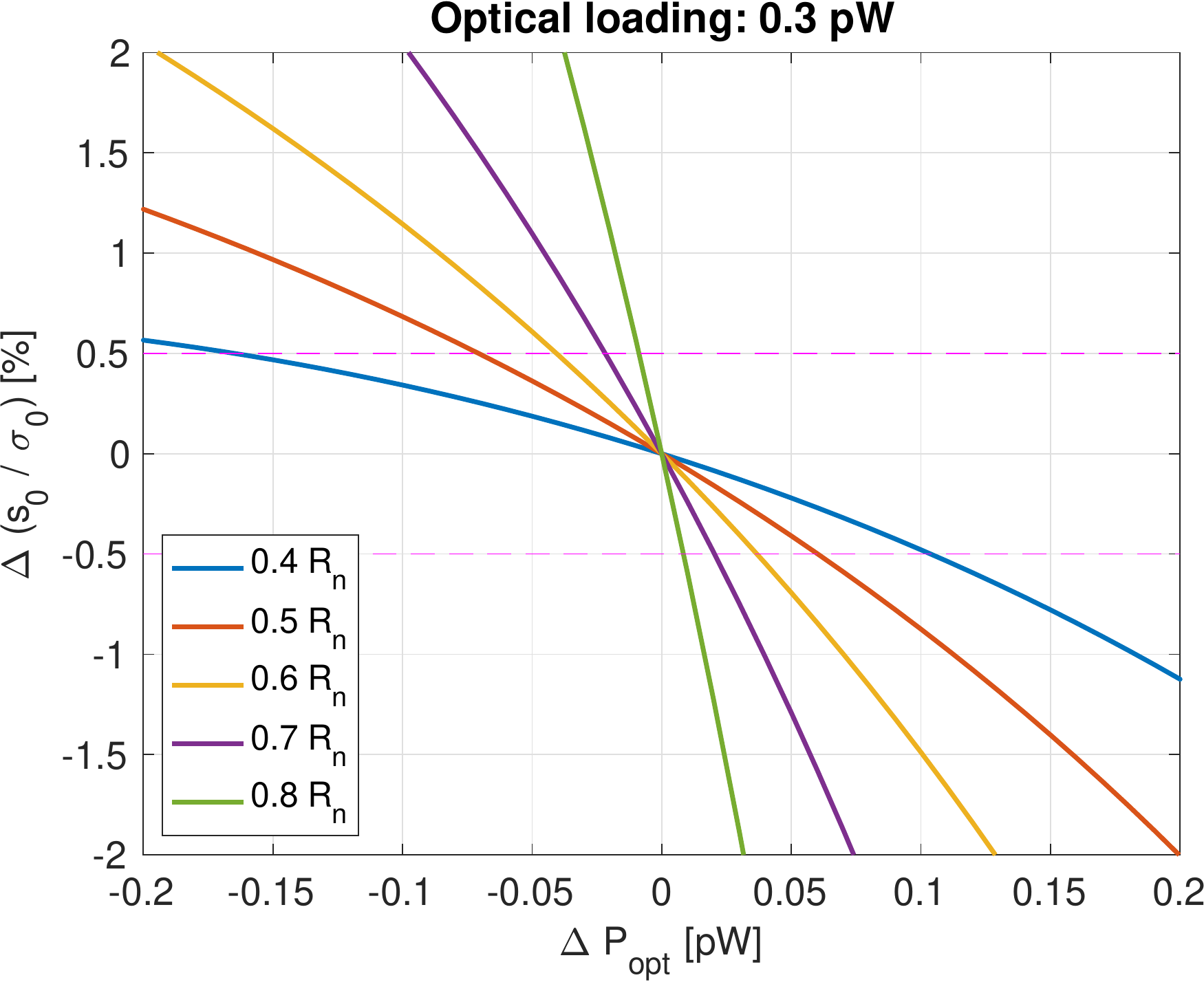}
    \caption{Representative simulation results for a \spider-like TES under flight loading. 
    \emph{Left:} Simulated fractional variation in zero-frequency optical responsivity, $s_0$, as a function of change in absorbed radiation power at fixed bias current. Different line colors indicate different initial TES resistances. Horizontal dotted lines indicate our target of $\pm0.5\%$ knowledge of responsivity.
    \emph{Right:} Similar plot for the variation in the ratio of the proposed estimator, $\sigma_0$, to the true responsivity, $s_0$. 
    }
   \label{fig:sims}
\end{figure}

Figure~\ref{fig:sims} shows representative results from these simulations, for different choices of initial TES resistance.
We find that changes in optical power $\lesssim 20$~fW (equivalent to wafer temperature changes of $\sim3$~mK) are sufficient to drive TES power responsivity outside of our $\pm0.5$\% target range.
Our proposed estimator, $\sigma_0$, tracks these variations well enough to keep the residual error within our target range under larger changes in incident power, but only at relatively low TES resistances (high loop gains). 
At bias points similar to those used in \spider's flight ($R\sim0.5 R_n$), this gain estimator improves our error on relative responsivity by a factor of $\sim3$, corresponding to $>10\%$ variation in total optical loading.
When the TES is nearly normal ($R\gtrsim0.8 R_n$, $\mathcal{L}_I \sim 8$), however, calibration with $\sigma_0$ performs slightly worse than simply assuming $s_0 \sim -\frac{1}{I_b R_{sh}}$.

\section{Application to \spider 2015}
\label{sec:flight}

During its 2015 flight, \spider applied bias step measurements at regular intervals throughout normal observations.
Bias steps were applied at the endpoint (``turnaround'') of every fifth azimuthal scan or every five minutes, whichever came first.
Each measurement consisted of a square wave applied to all TESs simultaneously atop the existing TES bias levels, with a frequency of 2~Hz, a duration of 2~s (4 complete cycles), and a peak-to-peak amplitude of 20 ADC bits.
This amplitude is $\sim$1\% of typical TES bias current, but sufficient to allow an amplitude measurement to $\leq 0.5\%$. 
Response to these bias steps was recorded as part of the regular TES time streams and analyzed locally in real-time by the flight operation system.
Since bias steps occurred only during scan turnarounds, they did not affect data used in science analysis.

Flight bias steps are used for two purposes: real-time monitoring of TES resistance, and post-flight characterizing of gain variations. 
The monitoring function adjusts the TES bias if the average resistance of detectors on that bias line has drifted out of a specified tolerance range.
The monitoring system can also trigger a bias adjustment if a specified fraction of detectors has become superconducting.
The detailed operations and algorithms are described in more detail in~\cite{RahlinThesis}.
During this first flight the re-biasing algorithm was fairly aggressive, adjusting bias for resistance changes $\geq0.02R_n$.
Note that this active feedback on the bias state complicates evaluation of the gain estimator's performance.

\begin{figure}
  \centering
  \includegraphics[width=0.6\linewidth, keepaspectratio]{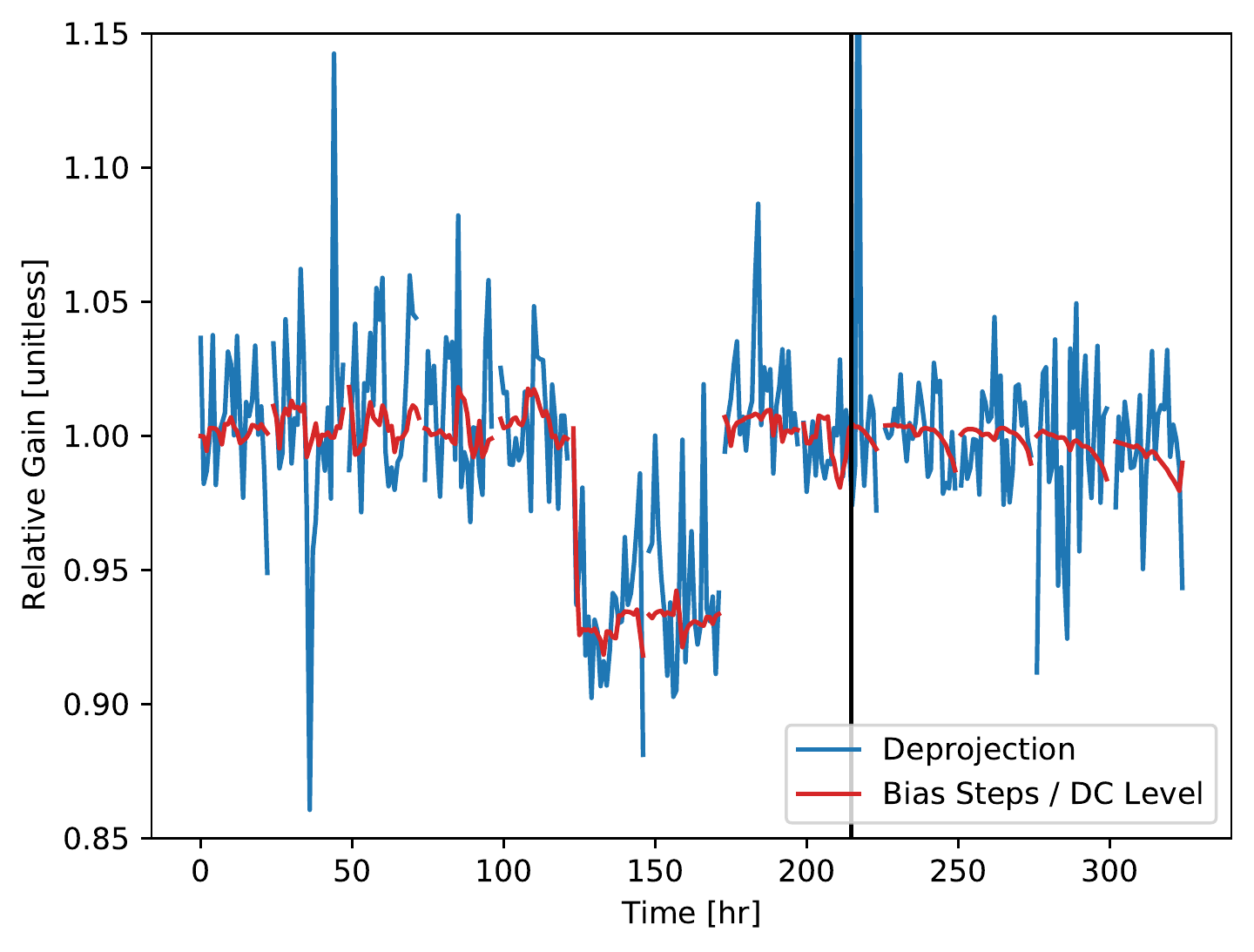}
  \caption{Comparison of relative detector gain over flight computed using different methods, averaged over a single focal plane. 
    In \emph{blue}, the gain is found through deprojection of a Planck frequency map.
    In \emph{red}, the relative gain is computed from bias steps and the DC level of the channels.
    The \emph{black} vertical line indicates where the transition from using bias steps to DC level for gain calibration occurs.
    The discrete change on flight days 6 and 7 is due to a temporary change in the targeted $R/R_n$.
    }
   \label{fig:gain_v_cal}
\end{figure}

Estimates of the detector responsivity from bias steps using Eq.~\ref{eqn:si0est} imply gain excursions on all time scales of $<5\%$, which we correct for using this estimator~\cite{Spider2021}.
We further find that these variations are not strongly correlated among detectors or with observational features.
Simulations show that, even if left uncorrected, the inferred gain variations would have had negligible effect on \spider's science observations~\cite{Spider2021}.
As an example, Figure~\ref{fig:gain_v_cal} shows inferred gain variations averaged across a single 95~GHz focal plane. The bias step estimate is generally consistent with the responsivity inferred from ``deprojection,'' in which 10-minute chunks of data are regressed against an appropriate \planck temperature map to provide a direct (if noisy) estimate of optical responsivity.
During the last few days of flight, bias steps (and active bias adjustment) were inadvertently disabled by a software bug; after this point we use an alternate gain proxy created by regressing the mean (``DC'') TES signal level against bias step amplitude from earlier in the flight.
The two estimators remain generally consistent during this period, though larger changes in response are visible than before due to the uncorrected effects of temperature drifts.

\section{Conclusions}
\label{sec:conclusion}

We have implemented a simple scheme for rapid, non-disruptive monitoring of TES bias state during observations using small electrical bias steps.
These can be used to derive a useful estimator for TES optical responsivity for TESs operated at high electrothermal feedback loop gain.
This technique was implemented successfully for \spider's 2015 balloon flight and may be a useful tool for calibration of future ground, balloon, and space instruments.

\begin{acknowledgements}
\spider is supported in the U.S. by the National Aeronautics and Space Administration under grants NNX07AL64G, NNX12AE95G, NNX17AC55G, and 80NSSC21K1986 issued through the Science Mission Directorate and by the National Science Foundation through PLR-1043515.
Logistical support for the Antarctic deployment and operations is provided by the NSF through the U.S. Antarctic Program.
Support in Canada is provided by the Natural Sciences and Engineering Research Council and the Canadian Space Agency.
Support in Norway is provided by the Research Council of Norway.
Support in Sweden is provided by the Swedish Research Council through the Oskar Klein Centre (Contract No.\ 638-2013-8993) as well as a grant from the Swedish Research Council (dnr.\ 2019-93959) and a grant from the Swedish Space Agency (dnr.\ 139/17).
The Dunlap Institute is funded through an endowment established by the David Dunlap family and the University of Toronto.
The multiplexing readout electronics were developed with support from the Canada Foundation for Innovation and the British Columbia Knowledge Development Fund.
KF holds the Jeff \& Gail Kodosky Endowed Chair at UT Austin and is grateful for that support.
WCJ acknowledges the generous support of the David and Lucile Packard Foundation, which has been crucial to the success of the project.
CRC was supported by UKRI Consolidated Grants, ST/P000762/1,
ST/N000838/1, and ST/T000791/1.
The data analyzed during the current study are available from the corresponding author on reasonable request.
\end{acknowledgements}

\pagebreak

\end{document}